\documentclass[a4paper,fleqn,usenatbib,useAMS]{mnras}
\usepackage{mathptmx}
\usepackage{graphicx}
\usepackage{amsmath}
\usepackage[utf8]{inputenc}  
\usepackage[T1]{fontenc}
\usepackage{amssymb}
\usepackage{wasysym}
\usepackage{epstopdf}
\usepackage[english]{babel}
\usepackage{multirow}
\usepackage{geometry}
\usepackage{hyperref}
\hypersetup{
    colorlinks=true,
    linkcolor=black,
    filecolor=magenta,      
    urlcolor=black,
    citecolor=black,
}

\title[Radio observations of a BCG at z=1.71]{Multiwavelength radio observations of a Brightest Cluster Galaxy at z=1.71: Detection of a modest Active Galactic Nucleus and evidence for extended star formation}

\author[A. Trudeau et al.]{Ariane Trudeau$^{1,2,3}$\thanks{Contact e-mail: \href{mailto:arianetrudeau@uvic.ca}{arianetrudeau@uvic.ca}}, \href{http://orcid.org/0000-0002-0104-9653}{Tracy Webb$^{2}$}, \href{http://orcid.org/0000-0001-7271-7340}{Julie Hlavacek-Larrondo$^{3}$}, \href{http://orcid.org/0000-0003-1832-4137}{Allison Noble$^{4}$},%, T. Webb, J. Hlavacek-Larrondo, A. Noble
\newauthor \href{http://orcid.org/0000-0002-7326-5793}{Marie-Lou Gendron-Marsolais$^{5}$}, Christopher Lidman$^{6}$, \href{http://orcid.org/0000-0003-4440-259X}{Mar Mezcua$^{7,8}$},% Adam Muzzin $^{9}$,
\newauthor Adam Muzzin$^{9}$, \href{http://orcid.org/0000-0002-6572-7089}{Gillian Wilson$^{10}$}, H. K. C. Yee$^{11}$
\\
$^{1}$Department of Physics \& Astronomy, University of Victoria, 3800 Finnerty Road, Victoria, British Columbia, V8W 2Y2, Canada\\%}% \\
$^{2}$Department of Physics, McGill University, 3600 rue University, Montréal, Québec, H3P 1T3, Canada\\%}
$^{3}$Département de Physique, Université de Montréal, Succ. Centre-Ville, Montréal, Québec, H3C 3J7, Canada\\
$^{4}$MIT Kavli Institute for Astrophysics \& Space Research, 77 Massachusetts Avenue, Cambridge, MA 02139, USA\\
$^{5}$European Southern Observatory, Alonso de Córdova 3107, Vitacura, Casilla 19001, Santiago, Chile\\
$^{6}$The Research School of Astronomy and Astrophysics, Australian National University, ACT 2601, Australia\\
$^{7}$Institute of Space Sciences (ICE, CSIC), Campus UAB, Carrer de Magrans, E-08193 Barcelona, Spain\\
$^{8}$Institut d'Estudis Espacials de Catalunya (IEEC), Carrer Gran Capità, E-08034 Barcelona, Spain\\
$^{9}$Department of Physics \& Astronomy, York University, 4700 Keele St., Toronto, Ontario, Canada, MJ3 1P3\\
$^{10}$Department of Physics \& Astronomy, University of California Riverside, 900 University Avenue, Riverside, CA 92521, USA\\
$^{11}$Department of Astronomy \& Astrophysics, University of Toronto, 50 St. George Street, Toronto, ON, M5S 3H4, Canada}
\date{Accepted 2019 May 14. Received 2019 May 3; in original form 2019 February 16}

\pubyear{2019}

\begin{document}
\label{firstpage}
\pagerange{\pageref{firstpage}--\pageref{lastpage}}
\maketitle
 %test

\begin{abstract}
 %***  \textbf{CHANGES ARE IN BOLD} ***

We present deep, multiwavelength radio observations of SpARCS104922.6+564032.5, a $z=1.71$ galaxy cluster with a starbusting core. Observations were made with the \textit{Karl G. Jansky Very Large Array} (JVLA) in 3 bands: 1-2 GHz, 4-8 GHz and 8-12 GHz. We detect a radio source coincident with the Brightest Cluster Galaxy (BCG) that has a spectral index of $\alpha$=0.44$\pm$0.29 and is indicative of emission from an Active Galactic Nucleus. The radio luminosity is consistent with the average luminosity of the lower redshift BCG sample, but the flux densities are 6$\sigma$ below the predicted values of the star-forming Spectral Energy Distribution based on far infrared data. Our new fit fails to simultaneously describe the far infrared and radio fluxes. This, coupled with the fact that no other bright source is detected in the vicinity of the BCG implies that the star formation region, traced by the infrared emission, is extended or clumpy and not located directly within the BCG. Thus, we suggest that the star-forming core might not be driven by a single major wet merger, but rather by several smaller galaxies stripped of their gas or by a displaced cooling flow, although more data are needed to confirm any of those scenarios. 

\end{abstract}

%\keywords{galaxies: active, galaxies: clusters: individual (SpARCS104922.6+564032.5), galaxies: evolution, galaxies: interactions, galaxies: starburst, radio continuum: galaxies}

% 6 keywords maximum, en ordre alphabétique

\section{Introduction}\label{Introduction}

The most massive galaxies in the Universe lie at the centre of galaxy clusters. In comparison with field galaxies, these objects, called Brightest Cluster Galaxies (BCGs), exhibit unique properties such as distinct luminosity and surface brightness profiles \citep[e.g.][]{oemler_structure_1976,tremaine_test_1977,dressler_comprehensive_1978}. We still do not understand how exactly BCGs acquired these distinct properties. Nevertheless, we suspect that environmental effects and their distinct formation histories might explain why they developed specific brightness profiles and luminosities.

On larger scales, the clusters in which BCGs reside can generally be divided into two categories: cool core clusters, which exhibit very peaked surface brightness distributions at X-ray wavelengths, and non cool core clusters, with similar overall X-ray luminosities but with smoother, less peaked X-ray surface brightness distributions. Some authors \citep[e.g.][]{hudson_what_2010,santos_evolution_2010} define an intermediate category called moderate or weak cool core clusters. Since cool core clusters have short radiative cooling time-scales on the order of $10^8 $ years in their centres \citep[e.g.][]{voigt_thermal_2004,mcnamara_heating_2007,mcnamara_mechanical_2012,hlavacek-larrondo_extreme_2012}, starbursts are expected to be common at the centre of such clusters. Indeed, the central cool gas in these clusters should condense onto the BCG, forming stars at rates of hundreds of solar masses per year \citep[e.g.][]{fabian_cooling_1994}. However, most BCGs are relatively quiescent and those that do show evidence of star formation generally tend to have star formation rates %(SFRs) 
one order of magnitude smaller, on the order of $1-150 ~\mathrm{M_{\odot} ~{yr}^{-1}}$ \citep[e.g.][]{donahue_infrared_2007,bildfell_resurrecting_2008,odea_infrared_2008,odea_hubble_2010,rawle_relation_2012}. %However, there are some exceptions; perhaps the most extreme of them is the BCG in the Phoenix cluster, with a SFR of $610\pm 50 ~\mathrm{M_{\odot}~{yr}^{-1}}$  \citep[e.g.][]{mcdonald_deep_2015,mcdonald_massive_2012,mcdonald_hst/wfc3-uvis_2013,mcdonald_state_2014,russell_alma_2017}.

This mismatch between expected and observed star-forming rates, known as the cooling flow problem, is thought to be caused by %was solved by taking into account the 
Active Galactic Nuclei (AGN) feedback processes from the BCG. AGNs can release copious amounts of energy into the intracluster medium (ICM) through many ways, including: jetted outflows that inflate cavities, weak shocks, sound waves or turbulence in the ICM \citep[e.g.][]{mcnamara_heating_2007,mcnamara_mechanical_2012,markevitch_shocks_2007,zhuravleva_turbulent_2014,fabian_sound_2017}. Alone, the energy released by jetted outflows appears to be on the same order as the energy needed to offset cooling \citep[e.g.][]{rafferty_feedback-regulated_2006,mcnamara_heating_2007,hlavacek-larrondo_extreme_2012}, therefore suggesting that AGN feedback is a good candidate for solving the cooling flow problem.

According to semi-analytic models \citep[e.g.][]{de_lucia_hierarchical_2007} and to several hydrodynamical simulations \citep[e.g.][]{ragone-figueroa_bcg_2018} it has been proposed that star formation occurs very early in BCG history (mostly before $z\sim 3$) and is quickly suppressed by AGN feedback \citep[e.g.][]{croton_many_2006}. Later, BCGs are thought to be built-up by dry mergers, without significant star formation. This scenario is supported by several mass growth measurements of BCGs, mostly below $z \sim 1$ \citep[e.g.][]{stott_near-infrared_2008,stott_little_2011,lidman_evidence_2012,bellstedt_evolution_2016}, although authors disagree on the mass growth rate. For example, \citet{stott_little_2011} measured a growth rate of 30\% between a redshift of $z=1$ and $z= 0.25$, while \citet{lidman_evidence_2012} found that BCG sizes increase by a factor of $1.8\pm 0.3$ between $z\sim 0.9$ and $z\sim 0.2$.

%on the epoch of BCGs growth. 

At $z>1$, however, there are growing divergences between this scenario and observations: \citet{webb_star_2015} and \citet{mcdonald_star-forming_2016} both find evidence of significant in-situ star formation in BCGs at $z\gtrsim 1$. We don't know what the star formation triggering mechanism is, but gas-rich galaxy interactions are a possibility. %but little evidence point toward gas-rich galaxy interactions}. 
For example, according to \citet{mcdonald_star-forming_2016}, star-forming BCGs seem to preferentially lie in dynamically unrelaxed, non-cool core clusters.

SpARCS104922.6+564032.5 (hereafter referred as SpARCS1049), located at $z=1.7089$ \citep{webb_extreme_2015}, provides additional evidence for this scenario. It is one of the most distant spectroscopically confirmed clusters known to date and was discovered by the \textit{Spitzer Adaptation of the Red-sequence Cluster Survey} (SpARCS) collaboration \citep[e.g.][]{muzzin_spectroscopic_2009,wilson_spectroscopic_2009,demarco_spectroscopic_2010}. 

The complex morphology of the BCG in SpARCS1049, revealed by the \textit{Hubble Space Telescope} (see Figure \ref{fig_radio}) suggests that this BCG has been caught in the process of a major merger \citep{webb_extreme_2015}. The single, large backward J-like tidal tail and the chain of clumps are reminiscent of \lq shrimp-like\rq~interacting galaxies, as defined by \citet{elmegreen_smooth_2007}. The clump chain seems to originate from within the stellar halo of the BCG and has a linear extent of $\sim 60$ kpc. 

The cluster core, including the BCG, a pair of interacting cluster members and another member (highlighted in Figure \ref{fig_radio}), is coincident with strong mid-/far infrared emission. Assuming all of the infrared flux could be attributed to the same object, \citet{webb_extreme_2015} measured an SED-fitted star formation rate of 860 $\mathrm{M{_\odot} ~yr^{-1}}$, after correcting for AGN contamination. However, the data used to calculate the star formation rate and build the spectral energy distribution (SED) suffer from poor spatial resolution (several to tens of arcsecs), and moreover, the centroid of the 24 $\umu$m \textit{Multiband Imaging Photometer} (MIPS) data used in this calculation is located approximately 15 kpc (1.75 arcsec) to the South-East of the BCG centre. Therefore, it is still unclear how extended, clumpy and where exactly is the star formation in the core of SpARCS1049. Hereafter, we will refer to this region of star formation as the star formation in the vicinity of the BCG.

Recently, a large reservoir of cold molecular gas ($\mathrm{M_{H_2}}=1.1\pm 0.1 \times 10^{11} ~\mathrm{M_{\odot}}$) was discovered in the central region of SpARCS1049 %\citep[\textbf{based on a source brightness temperature of $1.16\pm 0.10 \times 10^{11}  ~\mathrm{K~ km~ s^{-1}~ pc^{2}}$}]{webb_detection_2017}
%\citep[\textbf{based on a source brightness temperature of $1.16\pm 0.10 \times 10^{11}  ~\mathrm{K~ km~ s^{-1}~ pc^{2}}$ and a conservative $\alpha_{CO}$ of $0.8 ~\mathrm{M_\odot (K~ km~ s^{-1}~ pc^{2})^{-1}}$}][\textbf{. Choosing an alpha of $4.0~\mathrm{M_\odot (K~ km~ s^{-1}~ pc^{2})^{-1}}$, would have yield to a mass of $\mathrm{M_{H_2}}=5.5\pm 0.5 \times 10^{11} ~\mathrm{M_{\odot}}$}]{webb_detection_2017}.
\citep[based on a source brightness temperature of $1.16\pm 0.10 \times 10^{11} ~\mathrm{K~ km~ s^{-1}~ pc^{2}}$;][]{webb_detection_2017}. The CO-to-H2 conversion factor used is $\alpha_{CO}=0.8 ~\mathrm{M_\odot~ (K~ km~ s^{-1}~ pc^{2})^{-1}}$, providing a conservative estimate of the mass: $\alpha_{CO}$ could be as high as $4.0~\mathrm{M_\odot ~[K~ km~ s^{-1}~ pc^{2}]^{-1}}$ \citep{carilli_cool_2013}, yielding to a mass of $\mathrm{M_{H_2}}=5.5\pm 0.5 \times 10^{11} ~\mathrm{M_{\odot}}$. However, the beam, with a full width half-maximum of 25 arcsec, is too wide to constrain more precisely the gas location and extent. This gas could be fuelling the star formation through a galaxy merger with the BCG, but the lack of multiple velocity peaks, as might be expected in a major merger, as well as the immense amount of molecular gas, opens the door to other scenarios. For example, the core morphology and the gas reservoir could have been produced by the stripping of several smaller galaxies in the cluster centre. Another explanation %(M. Voit, private communication) 
is that a collision with an infalling galaxy disrupted the feedback mechanisms of the AGN in the BCG, triggering a cooling flow.

To explore the radio properties of this unique cluster and further constrain the location of the star-forming zone, we present deep, multiwavelength \textit{Karl G. Jansky Very Large Array} (JVLA) observations of SpARCS1049. Although a wide-field image has been obtained, we focus our analysis on the BCG and its vicinity. In Section \ref{observations}, we present the observations and data reduction of the JVLA datasets. In Section \ref{results} we analyze the radio data and in Section \ref{discussion} we discuss the results. Finally, we present a summary in Section \ref{summary}. Throughout this paper, we assume $H_0=69.6 ~\mathrm{km ~s^{-1} ~Mpc^{-1}}$, $\Omega_\mathrm{M}=0.286$ and $\Omega_\mathrm{\Lambda}=0.714$. At the redshift of the source \citep[$z=1.7089$,][]{webb_extreme_2015}, 1 arcsec corresponds to 8.610 kpc.

\section{Observations and data reduction}\label{observations}

\subsection{VLA observations and data reduction}\label{data_reduction}

\begin{table*}%{c c c c c c c}[htb!]
\caption{VLA observations.} 
\label{tab:table_test}
%%\tablenum{1}
\begin{tabular}{c c c c c c c}
\hline
Date & Frequency (Band) & Bandwidth & Configuration & On-source time & Flag percentage &RMS $^a$ \\ 
 & (GHz) & (GHz) &  & (min) & (\%) & $\umu\mathrm{Jy ~beam^{-1}}$ \\
\hline
%\hline
2016 Nov 19 & 1.5 (L) & 1 & A & 80 & 60 & 11\\
2016 May 20 \& 21 & 6 (C) & 4 & B & 54 & 45 & \multirow{2}{1em}{3$^b$}\\      
2016 May 21 & 6 (C) & 4 & B & 54 & 52\\
2016 May 21 & 10 (X) & 4 & B & 54 & 28 & 4\\
%%\enddata
\hline
%%\tablenotetext{a}{Local noise level.}
%%\tablenotetext{b}{Local noise level for the merged image at 6 GHz}
\multicolumn{7}{l}{$^a$ Local noise level.}\\
\multicolumn{7}{l}{$^b$ Local noise level for the merged image at 6 GHz.}\\
\end{tabular}
\end{table*}

In 2016, we were awarded 6.5 hours of observations on the \textit{Karl G. Jansky Very Large Array} 
(project 16A-283, PI Hlavachek-Larrondo). Observations, array configurations and on-source time are presented in Table \ref{tab:table_test}. Centred on the BCG in SpARCS1049, the observations consisted of 2 hours in L band (1-2 GHz), two observations of 1.5 hours each in C band (4-8 GHz) and 1.5 hours in X band (8-12 GHz). RMS noise, beam FWHMs (hereafter simply referred to as \textit{beams}) and position angles for the final reduced images are provided in Table \ref{tab:table_flux}, in the third, fourth and fifth column respectively.

Data reduction was performed with CASA \citep[Common Astronomy Software Application,][]{mcmullin_casa_2007} following the steps described below. Most of the data reduction was performed with CASA 4.7.2, but final image in the X band was made with version 5.1.2.

First, corrupted antennae listed in the operator logs were removed. Then, prior to the automatic RFI flagging procedure, data were pre-calibrated using the tasks \textsc{gaincal}, \textsc{bandpass} and \textsc{applycal}. For each antenna, we examined the amplitude versus frequency plot with \textsc{plotcal} and flagged any abnormally low or high visibilities. We then proceeded with automatic RFI excision using the \textsc{rflag} and \textsc{extend} modes of the task \textsc{flagdata}. The \textsc{tfcrop} mode was used on the most RFI affected spectral windows and further flagging was made with the \textsc{manual} mode. After calibration, target data were split.

Images were made with the task \textsc{clean}, using a W-projection algorithm (mode \textsc{widefield}) and 480 w-planes to correct the sky curvature across the field of view \citep{cornwell_non-coplanar_2008}. We used Briggs weighing and a robustness parameter of 0, although we tested other parameters in Section \ref{tests_image}. To ensure a sufficient sampling of the respective beams, we used a pixel size of 0.25 arcsec in L band, 0.20 arcsec in C bands and 0.15 arcsec in X band. For each band, the first clean was performed using the interactive mode, which allowed the creation of a customised cleaning mask. We then applied a self-calibration procedure to the initial image. This procedure consists of deriving phase corrections with \textsc{gaincal}, applying them to the data using \textsc{applycal} and then making a new image. We tested several time solution intervals and solving procedures, and varied the number of self-calibrations performed. The deepest images were obtained with one round of self-calibration and a T Jones solving procedure applied to an infinite time solution interval. The two datasets in C band were imaged, self-calibrated and then re-imaged separately. Then, they were merged using the task \textsc{clean}, self-calibrated and merged anew. We used 100 000 iterations for merging and final cleans, except in X band where the final clean was performed with 45 000 iterations.
The last column of Table \ref{tab:table_test} summarizes the local noise level reached in each image. Essentially, considering the exposure times and percentage of flagged data, we were able to reach the thermal noise in each of the images. 
The central region surrounding the BCG is shown in Figure \ref{fig_radio}.

\subsection{Additional imaging of the VLA datasets}\label{tests_image}

To further search for evidence of extended radio emission (which might be related to star formation associated with the BCG), we made additional images using different resolutions and Briggs parameters. First, we produced images with identical pixel sizes (0.25 arcsec) and beams for all frequencies (L, C and X bands). We set the parameter \textsc{restoringbeam} to be identical to the default L band beam, $1.27\times 0.85 ~\mathrm{arcsec^2}$ at an angle of $-81\deg$. We then compared the fluxes pixel by pixel. We found no traces of faint or diffuse emission beyond the detected point sources.

%Our second test consisted of degrading the resolution of the L band image, in order to better capture faint, extended emission. We applied the same imaging procedure as in Section \ref{data_reduction}, except for the last clean, where the resolution was degraded to 2 arcsec per pixel \textbf{(resulting in a beam of roughly $8 \times 8 ~\mathrm{arcsec^2}$ and a rms around 11 $\umu\mathrm{Jy ~{beam}^{-1}}$)}. We also made two other tests in L band, this time changing the Briggs robustness parameter. In one test, the robust parameter was set to $-2$, which is equivalent to a uniform weighting (i.e. provide a better resolution but less sensitivity). In the other case, the robustness parameter was set to $2$, equivalent to natural weighting (better sensitivity at the expense of the resolution). \textbf{In both cases, the RMS is 15 $\umu\mathrm{Jy ~{beam}^{-1}}$ and the beam keep to default}. None of these additional images show conclusive traces of extended emission beyond the BCG point source. This remains the case if we apply similar procedures to the C and X band images.

Our second test consisted of degrading the resolution of the L band image, in order to better capture faint, extended emission. We applied the same imaging procedure as in Section \ref{data_reduction}, except for the last clean, where the resolution was degraded to 2 arcsec per pixel. We also made two other tests in L band, this time changing the Briggs robustness parameter. In one test, the robust parameter was set to $-2$, which is equivalent to a uniform weighting (i.e. provide a better resolution but less sensitivity). In the other case, the robustness parameter was set to $2$, equivalent to natural weighting (better sensitivity at the expense of the resolution). None of these additional images show conclusive traces of extended emission beyond the BCG point source. This remains the case if we apply similar procedures to the C and X band images.

Finally, because we were able to reach a very low noise level in the C band merged image at 6 GHz, in addition to clearly detecting the BCG, we decided to split the two original datasets (those presented in Table \ref{tab:table_test}). Each dataset, spanning 4 GHz to 8 GHz, was divided into two sub-datasets: one spanning 4 to 6 GHz and another spanning 6 to 8 GHz. We merged datasets with identical frequencies following the procedure explained in the last section: separate self-calibration and imaging, then two imaging with both datasets with a self-calibration procedure intercalated. Two images, respectively centred at 5 GHz and 7 GHz, were hence created, providing two additional flux densities to constrain the BCG spectral index. Details of the BCG detection, noise and beam dimension of these images are given in Table \ref{tab:table_flux}. Signal-to-noise ratios in the L and X band are both below 6$\sigma$ and therefore unlikely to yield to a 3$\sigma$ BCG detection if split in two. Hence, we did not proceed with such a decomposition in these bands. From now on, we will specify the frequency when referring to a C band image, to avoid any confusion.

\section{Results and Analysis}\label{results}

\subsection{Source Detection and Characterization within the Vicinity of the BCG} \label{radio_flux}
\begin{figure*}%[!th]
%\gridline{\fig{IR_et_radio_minceur_BCGv2.eps}{1.0\textwidth}{}
%          }
%\gridline{\fig{BCG_contour_v4.eps}{1.0\textwidth}{}          
%          }
%\gridline{\fig{BCG_MIPs.eps}{1.0\textwidth}{}          
%          }
%\includegraphics[width=16.5cm]{BCG_MIPs.eps}
\includegraphics[width=16.5cm]{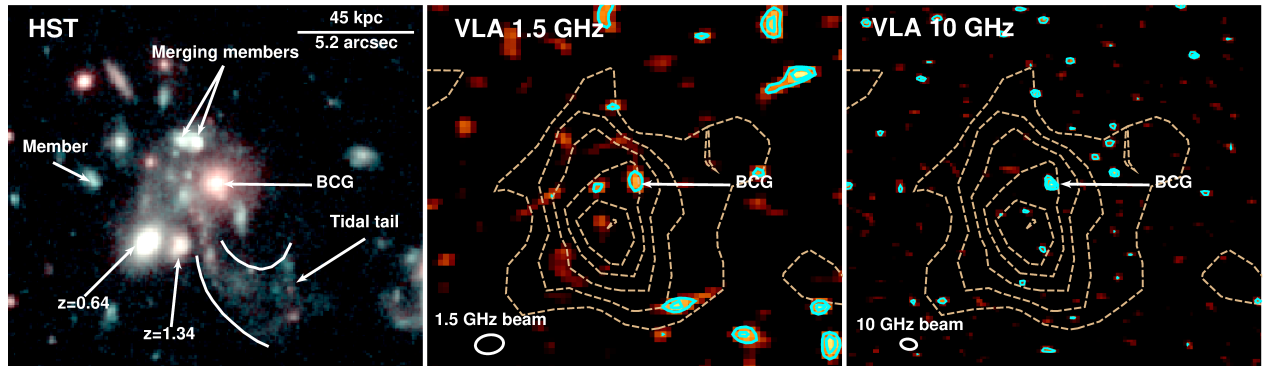}
\caption{Deep VLA images of the BCG in SpARCS1049 and its \textit{HST} counterpart.
Left: A composite \textit{HST} infrared image \citep[F105W in blue and green, F160W in red; for the data reduction see][]{webb_extreme_2015} %(F105W in blue and green, F160W in red)
of the centre of SpARCS1049. We highlight 3 cluster members, 2 foreground galaxies, the tidal tail and the BCG. %Zoom-in of Figure \ref{fig_centre_amas}.
Middle: 1.5 GHz VLA image of the same area ($\sigma_\mathrm{RMS}=10.5 ~\umu\mathrm{Jy ~{beam}^{-1}}$). Right: 10 GHz VLA image ($\sigma_\mathrm{RMS}=3.7 ~\umu\mathrm{Jy ~{beam}^{-1}}$). No extended emission is detected in any image. %Middle-left to middle-right: Same but for the second cluster member. The scale is indicated in the top right corner of the HST images. Bottom-left: Spitzer 4.5 $\mu$m image of the third member. Bottom-middle: L band image of the third member. Bottom-right: C band 7 GHz image of the third member (there is no detection of this member in X band), with a RMS of $4.7 ~\mu\mathrm{Jy ~{beam}^{-1}}$, $3\sigma_\mathrm{RMS}=14.1 ~\mu\mathrm{Jy ~{beam}^{-1}}$. 
The scale is indicated in the top right corner of the \textit{HST} image. The VLA beams are shown in the lower left corners. The radio contours (3, 4 and 5$\sigma_\mathrm{RMS}$ levels) are displayed in cyan and the \textit{Spitzer} MIPS 24 $\umu m$ contours are overplotted in beige.
}
\label{fig_radio}
\end{figure*}

We detected radio emission coincident with the BCG optical centre at $\gtrsim$ 3$\sigma_\mathrm{RMS}$ in each of our 5 radio images. The details of these detections are outlined in Table \ref{tab:table_flux} and shown in Figure \ref{fig_radio}. Although no other radio source is detected at or above 3$\sigma_\mathrm{RMS}$ in more than one band within 70 kpc (8 arcsec) of the BCG (to confirm a radio source, we required detections in at least two bands), we detected 2 spectroscopically confirmed clusters members as point sources. They lie East of the BCG at $10: 49: 26.40 +56: 40: 14.98$ and $ 10: 49: 32.25 +56: 40: 51.53$ (J2000).

Thus, within the BCG vicinity, the flux density of radio emission coincident with other optical sources does not exceed 3$\sigma_\mathrm{RMS}$ in two bands or more (see the second column of Table \ref{tab:table_radio_SFR}).

%\begin{deluxetable*}{c c c c c c c c c c}[htb!]
%\tablecaption{BCG detection \label{tab:table_flux}}
%%\tablenum{2}
%\center
%\tablehead{\colhead{Frequency} &\multicolumn{2}{c}{BCG} & \colhead{RMS} & \colhead{Beam} & \colhead{PA\tablenotemark{a}} \\ 
%\hline
%\colhead{} & \colhead{Integrated flux} &\colhead{Peak} & \colhead{} &\colhead{} \\
%\colhead{(GHz)} & \colhead{($\mu$Jy)} & \colhead{($\mu \mathrm{Jy ~{beam}^{-1}}$)}  &\colhead{($\mu\mathrm{Jy ~{beam}^{-1}}$)} & \colhead{(arcsec$\times$arcsec)} & \colhead{($\deg$)} } 
%\startdata
%\hline
%1.5 & $50\pm 21$ & $42\pm 11$  & 11 & $1.27\times 0.85$ & -81\\
%5 & $29.1\pm 9.8$ & $20.6\pm 4.1$ & 5 & $1.25\times 0.90$ & 40\\
%6 & $28.4\pm 8.7$ & $14.8\pm 3.0$  & 3 & $1.04\times 0.73$ & 38\\
%7 & $23.5\pm 8.7$ & $21.8\pm 4.7$  & 5 & $0.95\times 0.33$ & 37\\
%10 & $21.9\pm 6.5$ & $21.5\pm 3.6$ & 4 & $0.72\times 0.46$ & 81\\
%\enddata
%%\center
%\tablenotetext{a}{Position angle of the beam, measured counter-clockwise from North to East.}
%%\tablenotetext{b}{Upper limit on the flux of a potential point source, corresponding to 3 times the RMS of one beam}
%\end{deluxetable*}
%\renewcommand{\tabcolsep}{0.1cm}
%\renewcommand{\arraystretch}{0.75}
%\setlength{\arrayrulewidth}{1mm}
%\setlength{\tabcolsep}{1pt}
%\renewcommand{\arraystretch}{1.5}
%\renewcommand{\arraystretch}{0.8}
\begin{table*}%{c c c c c c c c c c}[htb!]
\caption{BCG detection.}
\label{tab:table_flux}
%\tablenum{2}
%\textwidth\tabcolsep
%\begin{tabular}{c c@{\hspace{-30ex}} c@{\hspace{-30ex}} c c c}
\begin{tabular}{cccccc}
%\begin{tabularx}{x x x x x x}
\hline
Frequency & Integrated flux & Peak & RMS & Beam FWHM & PA$^{a}$\\ %$^{a}$
%\hline
% & Integrated flux & Peak &  & \\
(GHz) & ($\umu$Jy) & ($\umu \mathrm{Jy ~{beam}^{-1}}$) & ($\umu\mathrm{Jy ~{beam}^{-1}}$) & (arcsec$\times$arcsec) & ($\deg$)\\%\umu \mathrm{Jy ~{beam}^{-1}}$
%\startdata
\hline
1.5 & $50\pm 21$ & $42\pm 11$ & 11 & $1.27\times 0.85$ & -81\\
5 & $29.1\pm 9.8$ & $20.6\pm 4.1$ & 5 & $1.25\times 0.90$ & 40\\
6 & $28.4\pm 8.7$ & $14.8\pm 3.0$ & 3 & $1.04\times 0.73$ & 38\\
7 & $23.5\pm 8.7$ & $21.8\pm 4.7$ & 5 & $0.95\times 0.33$ & 37\\
10 & $21.9\pm 6.5$ & $21.5\pm 3.6$ & 4 & $0.72\times 0.46$ & 81\\
\hline
%\enddata
%\center
\multicolumn{6}{l}{$^a$ Position angle of the beam, measured counter-clockwise from North to East.}\\
%\tablenotetext{a}{Position angle of the beam, measured counter-clockwise from North to East.}
\end{tabular}
%\end{tabularx}
\end{table*}

%\renewcommand{\tabcolsep}{0.2cm}
%\begin{deluxetable*}{c c c c c}[htb!]
%\tablecaption{SFR detection threshold for each image \label{tab:table_radio_SFR}}
%%\tablenum{3}
%\tablehead{\colhead{Frequency} & \colhead{3$\sigma_\mathrm{RMS}$ flux density upper limit} & \colhead{3$\sigma$ SFR upper limit} & \colhead{5$\sigma$ SFR upper limit} \\ 
%\colhead{(GHz)} & \colhead{($\mu\mathrm{Jy ~beam^{-1}}$)} & \colhead{($\mathrm{M_\odot ~yr^{-1} ~beam^{-1}}$)} & \colhead{($\mathrm{M_\odot ~yr^{-1} ~beam^{-1}}$)} } 
%\startdata
%\hline
%1.5 & 32 & 600 & 990 \\
%5 & 14 & 590 & 990\\%4.5*3=13.5
%6 & 10 & 480 & 800 \\
%7 & 14 & 780 & 1310 \\
%10 & 11 & 790 & 1320 \\
%\enddata
%\end{deluxetable*}

\begin{table*}%{c c c c c}[htb!]
\caption{Minimal star formation rate detectable for each image.}
 \label{tab:table_radio_SFR}
%\tablenum{3}
\begin{tabular}{c c c c}
\hline
Frequency & 3$\sigma_\mathrm{RMS}$ flux density upper limit & 3$\sigma$ SFR upper limit & 5$\sigma$ SFR upper limit \\ 
(GHz) & ($\umu\mathrm{Jy ~beam^{-1}}$) & ($\mathrm{M_\odot ~yr^{-1} ~beam^{-1}}$) & ($\mathrm{M_\odot ~yr^{-1} ~beam^{-1}}$) \\ 
%\startdata
\hline
1.5 & 32 & 600 & 990 \\
5 & 14 & 590 & 990\\%4.5*3=13.5
6 & 10 & 480 & 800 \\
7 & 14 & 780 & 1310 \\
10 & 11 & 790 & 1320 \\
\hline
%\enddata
\end{tabular}
\end{table*}

\begin{figure}
\centering
\includegraphics[width=5.7cm, angle=90]{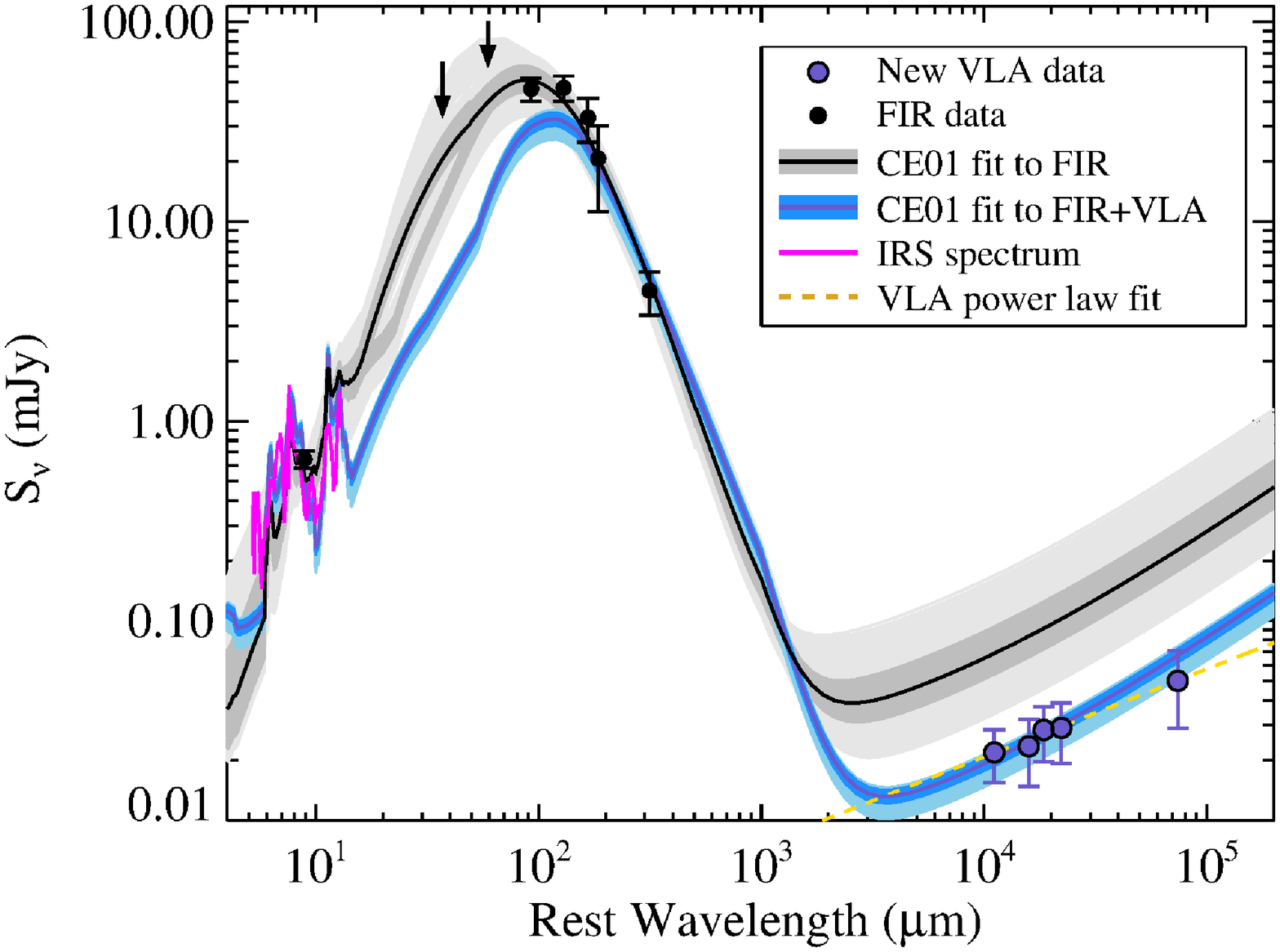}
%\caption{A comparison between our new JVLA data (in blue) and \citet{webb_extreme_2015} %Webb et al. (2015a) %\citet{webb_extreme_2015}
%best-fitting SED. The shaded region encloses all the SED variations within 1$\sigma$ (\textbf{grey and blue}) and 3$\sigma$ (light grey) of the best fit and the magenta line is the \textit{Infrared Spectrograph} (IRS) spectrum from \citet{webb_extreme_2015}. %Webb et al. (2015a). %\citet{webb_extreme_2015}. 
%In addition to the \textit{Spitzer} MIPS 24 $\umu$m data, the far-infrared data and limits are from the Photoconductor Array Camera and Spectrometer (PACS) and the Spectral and Photometric Imaging Receiver (SPIRE) onboard Herschel Space Observatory and from the Submillimetre Common-User Bolometer Array 2 (SCUBA-2) on James Clerk Maxwell Telescope. The dashed blue line shows the power law best fit to the radio point.The JVLA data have a reduced $\chi^2_\nu$ of 9.5 with respect to the SED best-fitting, which corresponds to an offset of approximately 6$\sigma$.
\caption{A comparison between \citet{webb_extreme_2015} best-fitting SED (in black) and the best infrared and radio fit (in blue). Both fits are based on \citet{chary_interpreting_2001} templates. The shaded region encloses all the SED variations within 68\% (medium grey and blue) and 95\% (lighter grey and blue) surface contour of the normalized $\chi^2$ probability distribution. The magenta line is the Infrared Spectrograph (IRS) spectrum from \citet{webb_extreme_2015}. In addition to the \textit{Spitzer} MIPS 24 $\umu$m data, the far-infrared data and limits are from the Photoconductor Array Camera and Spectrometer (PACS) and the Spectral and Photometric Imaging Receiver (SPIRE) onboard \textit{Herschel Space Observatory} and from the Submillimetre Common-User Bolometer Array 2 (SCUBA-2) on James Clerk Maxwell Telescope. The dashed yellow line shows the power law best fit to the radio point. There is an offset of approximately 6$\sigma$ between the VLA data and the original infrared-only fit.
}
\label{fig_SED}
\end{figure}

To understand the origin of the radio emission of the BCG, we analyzed its radio spectrum (see Figure \ref{fig_SED}). Although the point-like morphology of the detection and its coincidence with the optical centre of its \textit{HST} counterpart suggest that we detected an AGN, we do not exclude the possibility of a compact starburst. Indeed, compact starbursts are common in star-forming galaxies at $z\sim 2$ \citep[e.g.][]{elbaz_goods-herschel:_2011,barro_candels:_2013,elbaz_starbursts_2018}. In addition, several simulations of galaxy wet mergers \citep[e.g.][]{di_matteo_star_2007,hopkins_star_2013} found that the bulk of the post-merging starbursts occur at the centres of the newly formed galaxies.

Given that we have only five measured flux points we limit our radio SED fit to a simple power law:

\begin{equation}\label{eq_power_law}
S_\mathrm{model}=A\nu^\alpha
\end{equation}

where S is the flux density, A a coefficient and $\alpha$ the spectral index. This is appropriate for core or jet emission, as well as star formation, depending on the spectral index. The best fit for the BCG (see Figure \ref{fig_SED}) has a spectral index of -0.44, with a 1$\sigma$ uncertainty of 0.29. For the other members, we obtain spectral indexes of $-0.53\pm 0.19$ and $-0.76\pm 0.35$ respectively.

The spectrum of radio emission related to star formation tends to be steeper than our BCG best fit, with spectral indexes -0.8 $< \alpha <$ -0.7 \citep[e.g.][]{heesen_radio_2014}, but still overlap with its 1$\sigma$ confidence interval. Therefore, we cannot rule out the compact starburst hypothesis based solely on the spectral index. However, the 24 $\umu$m emission, used by \citet{webb_extreme_2015} as a tracer of the star formation, has a centroid that lies to the south-east of the BCG ($\sim 15$ kpc from the BCG optical centre), rather than on the BCG optical - and radio - centre. Additionally, we note that the radio flux detected from the BCG is inconsistent with the star-forming spectral energy distribution fit to the infrared emission (shown in Figure \ref{fig_SED}) at $\sim$ 6$\sigma$. Together, these points suggest that the radio emission inside the BCG is powered by an AGN.

\subsection{Star Formation Rate Limits}\label{SFR_limit_radio}
For each of our 5 radio images, the flux densities corresponding to 3 times the noise (derived in last section and displayed in the second column of Table \ref{tab:table_radio_SFR}) can be used to compute an upper limit to the star formation rate (SFR) per beam i.e. the threshold above which radio emission from star formation would have been detected (see Table \ref{tab:table_radio_SFR}). First, we must convert those flux densities into \textit{k}-corrected 1.4 GHz luminosities using \citep[e.g.][]{van_weeren_distant_2014,delhaize_vla-cosmos_2017}: 

%\hfill
%\newpage
%\clearpage
%\vspace{180pt}

\begin{equation}\label{L_1.4}
L_\mathrm{1.4GHz}=\frac{4\pi D_L^2}{(1+z)^{\alpha +1}} \left(\frac{1.4}{\nu_\mathrm{band}}\right)^{\alpha} S_\mathrm{band}
\end{equation}

where $L_\mathrm{1.4GHz}$ is in $\mathrm{W ~Hz^{-1}}$, $D_\mathrm{L}$ is the luminosity distance, $\alpha$ the spectral index, $\nu_\mathrm{band}$ is the image frequency in GHz and $S_\mathrm{band}$ is the flux density in $\mathrm{W ~Hz^{-1} ~m^{-2}}$. Here, we assume a spectral index of $-0.7$, typical for star-forming regions \citep[e.g.][]{heesen_radio_2014}.

These 3$\sigma_\mathrm{RMS}$ upper limits of the \textit{k}-corrected 1.4 GHz luminosities can now be used to compute upper limits on the star formation rate within one beam, using the relation of \citet{condon_radio_2002}:

\begin{equation}\label{SFR}
SFR=1.20\times 10^{-21}L_\mathrm{1.4GHz}
\end{equation}

where the star formation rate (SFR) is in $\mathrm{M_\odot ~yr^{-1}}$ and $L_\mathrm{1.4GHz}$ is, as before, in $\mathrm{W ~Hz^{-1}}$. We used a Salpeter initial mass function to compute the conversion factor. Results are presented in the third column of Table \ref{tab:table_radio_SFR}, which also displays the results of a similar computation, starting this time with radio flux densities equivalent to 5$\sigma_\mathrm{RMS}$. The C band image at 6 GHz places the lowest constraint on the star formation rate within one beam (480 $\mathrm{M_\odot ~yr^{-1}}$ $\mathrm{~beam^{-1}}$; Table \ref{tab:table_radio_SFR}, column 3), but, to confirm a radio source, we required detections in at least two bands. Hence, we will consider the SFR upper limit to be 600 $\mathrm{M_\odot ~yr^{-1}}$ $\mathrm{~beam^{-1}}$ (L band detection threshold).

\section{Discussion}\label{discussion}

\subsection{Comparison to other BCGs}\label{comparison_BCG}

\begin{figure}
\centering
\includegraphics[width=8cm]{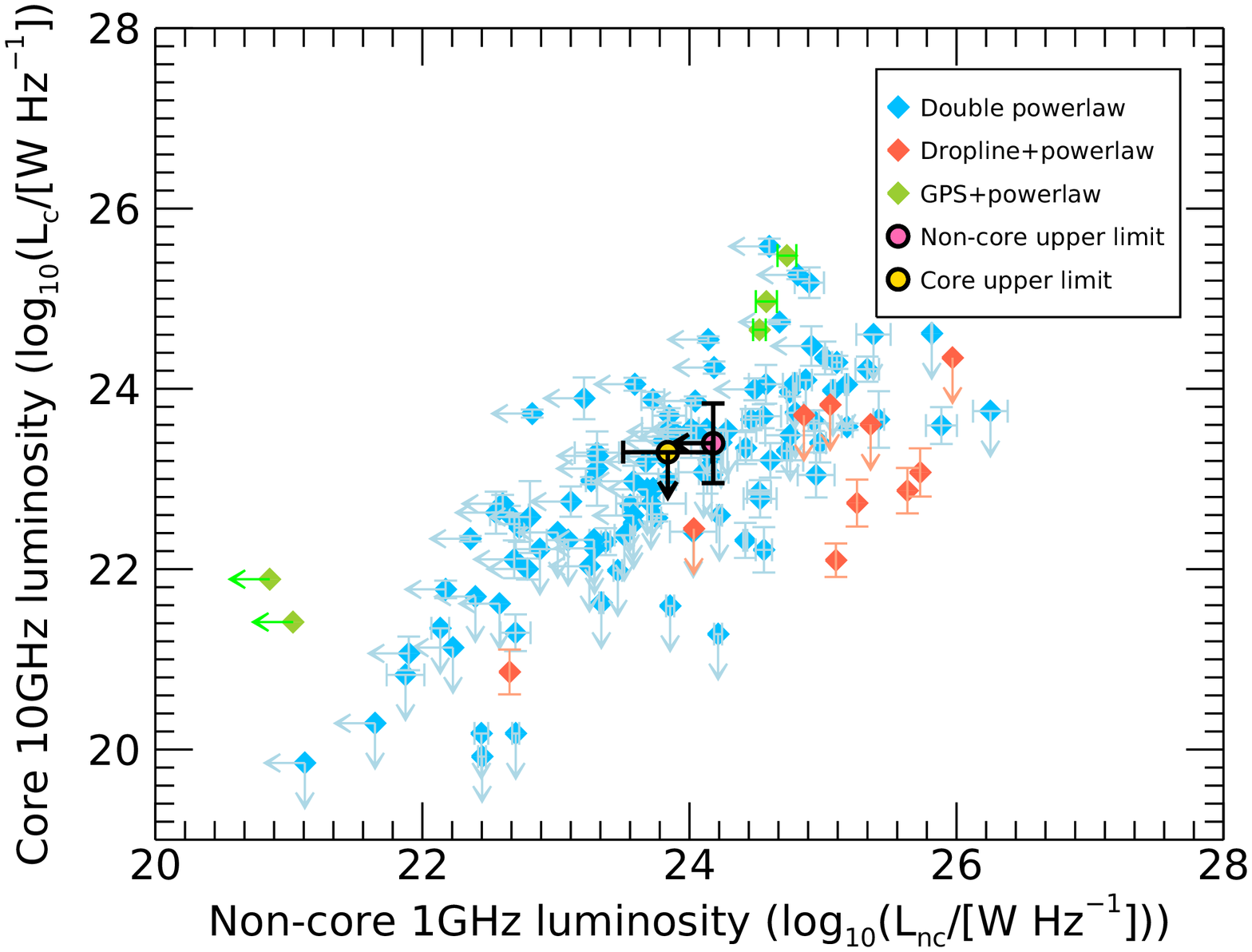}
\caption{A comparison between core and a non-core powers of the \citet{hogan_comprehensive_2015} %Hogan et al. (2015) %\citet{hogan_comprehensive_2015} 
sample and 2 possibilities for SpARCS1049. Blue diamonds are \citet{hogan_comprehensive_2015} %Hogan et al. (2015) %\citet{hogan_comprehensive_2015} 
objects following a power law model for non-core and core emission and coral diamonds non-core emission follow a \lq dropline model\rq , associated with a power law model for core emission. Green diamond models for non-core emission are power laws, but their core emission follow gigahertz peaked source (GPS) models. Assuming that most of the emission from SpARCS1049 BCG originates from one component, we derived an upper limit for the non-core component (pink circle) and the core component (gold circle).}
\label{fig_Hogan}
\end{figure}

We compare the radio emission from the BCG in SpARCS1049 with the radio luminosities of the lower redshift BCGs from the \citet{hogan_comprehensive_2015} line-emitting sample (mean redshift $z=0.12$). They determined the properties of about 250 $z\lesssim 0.4$ BCGs, half of them line-emitting, to explore if and how the radio properties of BCGs in relaxed clusters differ from those of BCGs in more disturbed clusters. These authors use the presence or absence of emission lines (especially H$\alpha$ and [NII]) as a proxy to distinguish between clusters with strong cool cores and those with weak or no cool cores. Given the intense star formation detected in SpARCS1049 BCG \citep{webb_extreme_2015}, these criteria would imply the presence of a cool core in this BCG (see Rhea et al., in preparation for a discussion of this source's X-ray properties). 

Since the steep, non core component usually dominates the overall BCG radio emission at low frequencies, \citet{hogan_comprehensive_2015} use the 1 GHz emission to characterize the non-core component. Conversely, the flat, core component usually dominates the BCG radio emission at higher frequencies. \citet{hogan_comprehensive_2015} hence use the 10 GHz emission to characterize the core component. Figure \ref{fig_Hogan} presents the core and non core luminosities of the line-emitting BCGs in \citet{hogan_comprehensive_2015}. The colours and shapes of the symbols indicate the various models used by \citet{hogan_comprehensive_2015} to perform the luminosity decompositions. 

We are unable to perform a detailed luminosity decomposition of the radio emission in SpARCS1049 because of the lack of spectral coverage. We therefore test two limiting cases. We first assume that all the emission we observed in the BCG comes from AGN jets and follow the spectral index computed in Section 3.1. Based on the emission level at 1.5 GHz in the observer frame, we computed the k-corrected jet emission at 1 GHz. Then based on the k-corrected jet emission at 10 GHz, we compute an upper limit for the core emission. The result is shown in gold on Figure \ref{fig_Hogan}. We then assume that all the observed radio emission in the BCG comes from the AGN core. Following a similar procedure, we compute a k-corrected core emission at 10 GHz and an upper limit for the jet emission at 1 GHz. The result is shown in pink on Figure \ref{fig_Hogan}. 

Figure \ref{fig_Hogan} shows that the BCG in SpARCS1049 lies in the middle of the dot cloud, no matter how the emission is distributed between jets and core. Therefore, the BCG in SpARCS1049 has a radio luminosity typical of BCGs at $z \leq 0.4$, which is somewhat surprising. Since SpARCS1049 hosts a $\sim 10^{11} ~\mathrm{M_\odot}$ molecular gas reservoir \citep{webb_detection_2017}, and radio AGN tend to be more common and more luminous at $z\sim 1.5$ \citep{smolcic_vla-cosmos_2017}, we expected the BCG of SpARCS1049 to lie among the luminous BCGs of the \citet{hogan_comprehensive_2015} sample.

%More quantitatively, assuming the simple power law model computed in Section \ref{radio_flux}, the BCG in SpARCS1049 has a rest frame 1.5 GHz luminosity of $5.8\pm 2.9 \times 10^{23} \mathrm{~W~Hz^{-1}}$, slightly below the threshold of $10^{24}~\mathrm{W~Hz^{-1}}$ set by \citet{best_cosmic_2014} to distinguish between luminous and less luminous radio-mode AGN.

%More quantitatively, \textbf{we can compare SpARCS1049 BCG to the 1.4 GHz luminosity function. Using the 1.4 GHz rest-frame characteristic luminosity computed by \citet{smolcic_vla-cosmos_2017}, for AGN with luminosities between $10^{22}$ and $10^{24}~\mathrm{W ~Hz^{-1}}$, we have}:
%
%\begin{equation}
% L^*(z) \propto (z+1)^{(2.88\pm 0.82)-(0.84\pm 0.34)z}
%\end{equation}
%
%\textbf{with $L^*(0)=10^{24.59}~\mathrm{W~Hz^{-1}}$. This yields to $L^*(1.71)=(1.64\pm 0.51)\times 10^{25}~\mathrm{W~Hz^{-1}}$. Assuming} the simple power law model computed in Section \ref{radio_flux}, the BCG in SpARCS1049 has a rest frame \textbf{1.4 GHz luminosity of $6.4\pm 3.0 \times 10^{23} \mathrm{~W~Hz^{-1}}$, less than an order of magnitude below the characteristic luminosity of \citet{smolcic_vla-cosmos_2017}.}

More quantitatively, we can compare the SpARCS1049 BCG to the 1.4 GHz luminosity function. \citet{smolcic_vla-cosmos_2017} calculated that the 1.4 GHz rest-frame characteristic luminosity for AGN with luminosities between $10^{22}$ and $10^{24}~\mathrm{W ~Hz^{-1}}$ is $ L^*(z) \propto (z+1)^{(2.88\pm 0.82)-(0.84\pm 0.34)z}$ with $L^*=10^{24.59}~\mathrm{W~Hz^{-1}}$ locally. This yields to $L^*=(1.64\pm 0.51)\times 10^{25}~\mathrm{W~Hz^{-1}}$ at the cluster redshift. Assuming the simple power law model computed in Section \ref{radio_flux}, the BCG has a rest frame 1.4 GHz luminosity of $6.4\pm 3.0 \times 10^{23} \mathrm{~W~Hz^{-1}}$, more than an order of magnitude below the characteristic luminosity of \citet{smolcic_vla-cosmos_2017}.

Therefore, the black hole at the centre of the BCG must have a modest accretion rate. This relatively low luminosity suggests that something may prevent or partially block inflows from the molecular gas reservoir. Among the possible culprits: an efficient star formation or an offset between the gas reservoir and the BCG.

\subsection{Star formation in the vicinity of the BCG}\label{SFR_discu}

\begin{table*}%{c c c c c c c}[htb!]
\caption{Summary of the best-fitting SEDs.} 
\label{tab:table_fits}
%%\tablenum{1}
\begin{tabular}{c c c c c c c}
\hline
Fit & $\chi^2_\nu$ & 60$\umu$m flux & Uncorrected SFR & SFR$^a$ \\ 
 & & (mJy) & ($\mathrm{M_\odot ~yr^{-1}}$) & ($\mathrm{M_\odot ~yr^{-1}}$) \\
\hline
%\hline
IR \& radio & 1.86 & $13\pm 1$ & $467\pm 29$ & $374\pm 23$ \\
IR only & 0.44 & $40^{+9}_{-6}$ & $1074\pm 162$ & $859\pm 130$ \\
%%\enddata
\hline
%%\tablenotetext{a}{Local noise level.}
%%\tablenotetext{b}{Local noise level for the merged image at 6 GHz}
\multicolumn{5}{l}{$^a$ Assuming a 20\% AGN contribution.}\\
\end{tabular}
\end{table*}
 
After a tentative characterization of the detected AGN, the obvious question is: where is the intense star formation indicated by the infrared emission? \citet{webb_extreme_2015} computed an AGN-corrected star-forming rate of $860\pm 130~\mathrm{M_\odot ~yr^{-1}}$, based on the far infrared emission (FIR), computed from the SED fitting.
 A comparison between our VLA and the \citet{webb_extreme_2015} spectral energy distributions is displayed in Figure \ref{fig_SED} and summarized in Table \ref{tab:table_fits}. This Figure reveals that the measured radio fluxes are significantly below \citet{webb_extreme_2015} SED fit. A computation of the reduced $\chi^2$ ($\chi^2_\nu$) of the radio data with respect to the best-fitting SED gives $\chi^2_\nu= 9.5$, which correspond to an offset of 6$\sigma$. Also, the expected spectral indexes, both around -0.7, are slightly steeper than the observed index, $-0.44\pm 0.29$. 
 
To investigate whether or not a different SED template could better fit the radio data, we performed our own fit of far-infrared and radio data, following \citet{webb_extreme_2015} method \citep[also described in][]{noble_phase_2016}: we tested each \citet{chary_interpreting_2001} templates over a range of amplitudes, generating a 2D-grid of fits. We select the best fit based on its $\chi^2$ probability. In Figure \ref{fig_SED}, the dark shaded regions correspond to all fits enclosed within the 68\% surface contour of the normalized $\chi^2$ probability distribution and the lighter shades correspond to all fits within 95\%. The reduced $\chi^2_\nu$ of the best fit is 1.86, corresponding to an integrated $\chi^2$ probability value of 0.05. Hence, we can reject the possibility that the fit accurately described the data with 95\% confidence. By contrast, with a $\chi^2_\nu$ of 0.44, the infrared-only fit can be rejected with 20\% confidence only. It is worth noticing that the infrared and radio fit provides a better estimation of the radio flux, although slightly too steep to match with the best-fitting power law. %(as discussed in Section \ref{radio_flux}, radio emission related to star formation tend to have a steeper slope than our power law fit). 
However, it significantly underestimates the far infrared emission: the 60 $\umu$m flux derived from the infrared-only fit is 3 times greater than the flux estimated from the radio and infrared fit (see Table \ref{tab:table_fits}). Therefore, no template seems to be consistent with both far infrared and radio emission.

This mismatch suggests that far infrared emission is tracing a different phenomenon than the radio. Since the radio emission probably comes from the AGN (see Section \ref{radio_flux}), the far infrared emission might be dominated by star formation. 
%One explanation for the discrepancy between \textbf{infrared and radio data is that \cite{chary_interpreting_2001} templates} assumes a star-formation dominated emission. However, as discussed in \textbf{Section\ref{radio_flux}}, the radio emission probably comes from the AGN. Therefore, we can treat this difference as an additional indication that the radio emission is AGN dominated with few (if any) contribution from the star formation.
Moreover, most of the mid and far-infrared measurements have resolutions spawning from several to tens of arcsecs, while our biggest radio beam spans 1.27 arcsec $\times$ 0.85 arcsec. Keeping this in mind, we suggest that the star formation happens close to the BCG but not within it. If the star formation had been within the BCG, we would have detected it (the BCG is a point source) and the new SED fit would have been more consistent with the far infrared data.
\begin{figure}
\centering
\includegraphics[width=8cm]{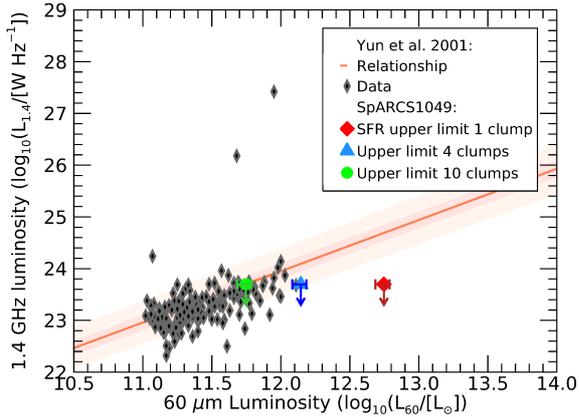}
\caption{Comparison between \citet{yun_radio_2001} (the 1 and 3$\sigma$ uncertainties are displayed as coloured regions) %Yun et al. (2001)%\citet{yun_radio_2001} 
and our upper limit for star formation detection, as calculated in Section 3.2. %\ref{SFR_limit_radio}. 
The upper limits in red blue and green show what happens when we vary the number of star formation \lq clumps\rq , based on the assumption that these clumps could be resolved in radio (so the detection limit does not change in radio), but are all within one beam in the infrared. We also assume that all \lq clumps\rq~contribute equally to the $60~\umu$m emission.
% the maximum radio luminosity of star formation in one beam, as calculated in Section \ref{upper_limit_radio}. 
The $60~\umu$m luminosity calculation is based on the best-fitting infrared SED. To be consistent with our SED-based star-forming rates, the AGN far-infrared luminosity is subtracted from the FIR luminosities presented here, assuming a 20 per cent contribution \citep{webb_extreme_2015}. %(Webb et al. 2015a).%\citep{webb_extreme_2015}. 
%The maximum radio luminosity is $2.6\sigma$ below the expected luminosity.%as for SED-based star formation rate 
%MESSAGE: La SF ne peux pas se concentrer dans un beam: elle est diffuse.
}
\label{fig_IR}
\end{figure}

To explore the extent or possible \lq clumpiness\rq~of this star-forming zone, we used the \citet{yun_radio_2001} relation between far infrared (60$~\umu$m) and 1.4 GHz luminosity. Figure \ref{fig_IR} shows the \citet{yun_radio_2001} data and relationship, with 3 overplotted upper limits corresponding to 1, 4 and 10 star-forming \lq clumps\rq . We assume that each \lq clump\rq~contributes equally to the FIR emission, while in radio they could be resolved. We set the radio flux upper limit to $5.0\times 10^{23}~\mathrm{W~Hz^{-1}}$ ($600~\mathrm{M_\odot ~yr^{-1}}$), in agreement with the L band detection threshold (see Section \ref{SFR_limit_radio}).

The star-forming region is likely to be clumpy, since the 10 clumps upper limit lies on the relationship while the 4 clumps upper limit is on the edge of its $3\sigma$ region. This is consistent with a non-detection: with a star formation rate of $860 \pm 130~\mathrm{M_\odot~ yr^{-1}}$, the star-forming zone is not detectable as long as it spreads over two VLA radio beams or more. This would be consistent with the star formation being distributed across the \lq beads on a string\rq~optical feature. We don't expect the star formation to be spread across the cluster galaxies in the beam: excluding the BCG, there is only 3 cluster members within the MIPs beam, none of them situated at less than $\sim 30$ kpc ($\sim 3.5$ arcsec) of the detection centroid (see Figure \ref{fig_radio}).

\subsection{Implications for BCG formation scenarios}\label{formation_scenario}
%Although the \textit{HST} mosaic revealed a morphology consistent with a merger-induced starburst \citep{webb_extreme_2015}, this scenario is disfavoured by our constraints on star formation. \citet{di_matteo_star_2007} and \citet{hopkins_star_2013} found that most of the post-merger starburst occurs in the centre of the newly formed galaxy. In SpARCS1049, the bulk of the star formation is likely occurring outside of the BCG. Moreover, the AGN in the BCG is not very active, which is weak evidence against this scenario: major galaxy interactions can trigger radio-mode AGN \citep{ellison_galaxy_2015}, but they are not the main cause of AGN activity \citep{kocevski_candels:_2012}. Besides, the gas reservoir in SpARCS1049 seems too large to originate from gas-rich mergers \citep[e.g.][]{edge_detection_2001,mcdonald_hst/wfc3-uvis_2013} and lacks the multiple velocity components induced by such events \citep{webb_detection_2017}. 

Although the \textit{HST} mosaic revealed a morphology consistent with a merger-induced starburst \citep{webb_extreme_2015}, this scenario is disfavoured by our constraints on star formation. \citet{di_matteo_star_2007} and \citet{hopkins_star_2013} found that most of the post-merger starburst occurs in the centre of the newly formed galaxy. In SpARCS1049, the bulk of the star formation is likely occurring outside of the BCG. Moreover, the AGN in the BCG is not very active, which is weak evidence against this scenario: major galaxy interactions can trigger radio-mode AGN \citep{ellison_galaxy_2015}. However, they are not the main causes of AGN activity, as only $16.7^{+5.3}_{-3.5}\%$ of the radio-AGN hosts are highly disturbed, a fraction consistent with the amount of non-active interacting galaxies \citep{kocevski_candels:_2012}. Besides, the gas reservoir in SpARCS1049 seems too large to originate from gas-rich mergers \citep[e.g.][]{edge_detection_2001,mcdonald_hst/wfc3-uvis_2013} and lacks the multiple velocity components induced by such events \citep{webb_detection_2017}. 

To explain the presence of a $1.1\pm 0.1\times 10^{11} ~\mathrm{M_\odot}$ gas reservoir in the centre of SpARCS1049, \citet{webb_detection_2017} suggested two other scenarios: 1) several smaller galaxies stripped of their gas by the BCG or 2) a cooling flow. While the first scenario cannot be rejected or confirmed by our radio data, the second scenario raise the possibility of a high-redshift Phoenix cluster analogue. 

The Phoenix cluster hosts the only other well-studied high redshift BCG with a comparable star formation rate. At z=0.596, its BCG hosts a massive cooling flow of $3820\pm 530~\mathrm{M_\odot ~yr^{-1}}$ fuelled by a molecular gas reservoir (dihydrogen) of $2.1\pm 0.3\times 10^{10} ~\mathrm{M_\odot}$ and forms stars at a rate of $610\pm 50~\mathrm{M_\odot ~yr^{-1}}$ \citep[see][]{mcdonald_massive_2012,mcdonald_hst/wfc3-uvis_2013,mcdonald_state_2014,mcdonald_deep_2015,ueda_suzaku_2013,tozzi_new_2015,russell_alma_2017}. By comparison, SpARCS1049 has a SFR of $860\pm 130~\mathrm{M_\odot ~yr^{-1}}$ and a $\sim 5$ times more massive molecular gas reservoir.

Despite such a large gas reservoir to fuel AGN feedback, the total 10 MHz to 10 GHz integrated power of SpARCS1049 central AGN is about $4.4\pm 3.5\times 10^{33}$ W. In contrast, the radio power emitted by the jets in the Phoenix cluster is $\sim 1000$ times stronger \citep[$3.6\times 10^{36}$ W;][]{mcdonald_deep_2015}. This suggests that, if there is a substantial cooling flow in the centre of SpARCS1049, only a small fraction of the inflowing gas reaches the centre of the BCG in SpARCS1049. Thus, if SpARCS1049 has a cooling flow, there might be a significant offset between the BCG and the gravitational centre of the cluster. \citet{hamer_relation_2012} and \citet{vantyghem_enormous_2019} %\citet{vantyghem_enormous_2018} 
suggested that the sloshing motion due to a galaxy interaction can trigger gas condensation offsetted from the BCG. A similar history could explain the presence of molecular gas in the core of SpARCS1049, although \citet{vantyghem_enormous_2019} %\citet{vantyghem_enormous_2018} 
argue that the condensation of all the intracluster medium within 10 kpc of the centre of the cluster is required to create a $10^{10}~\mathrm{M_\odot}$ gas reservoir in RXJ0821+0752. Therefore, additional mechanisms might be needed to explain the gathering of $10^{11}~\mathrm{M_\odot}$ of cold gas in SpARCS1049. Alternatively, the AGN weakness may originate from a very efficient star formation in SpARCS1049.

%Although the optical morphology of the Phoenix cluster BCG is filamentary \citep{mcdonald_deep_2015}, the majority of the star formation occurs within 15 kpc of the BCG centre (M. McDonald, private communication). If SpARCS1049 star-forming region has a similar extent, then the size of the star-forming region would be consistent with a non-detection in our VLA data. 

%Whether or not the offset between the BCG and the star forming region has been triggered by sloshing motion, the cooling flow hypothesis provides a compelling scenario to explain the peculiarities of SpARCS1049. However, X-ray data are clearly needed to distinguish between this scenario and gas stripping of multiple small galaxies.

\section{Summary}\label{summary}

We presented deep, multiwavelength JVLA radio observations of SpARCS1049, one of the most distant galaxy clusters ever studied in radio. SpARCS1049 is a cluster of galaxies with a starbursting core, displaying a complex morphology in the infrared \citep{webb_extreme_2015} and a immense $\sim 10^{11} \mathrm{M_\odot}$ molecular gas reservoir \citep{webb_detection_2017}. We detected the BCG at 1.5, 5, 6, 7 and 10 GHz, but we did not detect any diffuse emission or starbursting clump outside of the core. Given that, we draw the following conclusions:

%\begin{itemize}
%
%\item{The radio emission of the BCG likely comes from its AGN. It is best fit by a simple power law, with a spectral index of $-0.44\pm 0.29$ and has a radio luminosity consistent with the average luminosity of the lower redshift BCG sample of \citet{hogan_comprehensive_2015}. }
%\item{We found an offset of 6$\sigma$ between our new VLA data and the SED fit previously made \citep{webb_extreme_2015}, which suggests that star-forming regions are not embedded in the BCG. Combining the SED fit with our upper limit for the radio emission induced by star formation, we find that star forming regions are either made of numerous clumps or very extended.}
%\item{We explore the possible origins of the BCG complex infrared morphology and of its gas reservoir. There is growing evidence against the single, major wet merger scenario initially developed by \citet{webb_extreme_2015}. Although a cooling flow fits well with our constraints on the star-forming regions, X-ray data are needed to distinguish between this scenario and the gas stripping of multiples galaxies.}
%
%\end{itemize}
%
\begin{description}

\item{The radio emission of the BCG likely comes from its AGN. It is best fit by a simple power law, with a spectral index of $-0.44\pm 0.29$ and has a radio luminosity consistent with the average luminosity of the lower redshift BCG sample of \citet{hogan_comprehensive_2015}. }
\item{We found an offset of 6$\sigma$ between our new VLA data and the SED fit previously made \citep{webb_extreme_2015}. Moreover, our best infrared \& radio fit underestimates the far infrared fluxes and can be rejected with 95\% confidence based on the integrated $\chi^2$ probability. This suggests that star-forming regions are not embedded in the BCG. Combining the SED fit with our upper limit for the radio emission induced by star formation, we find that star-forming regions are either made of numerous clumps or very extended.}
\item{We explore the possible origins of the BCG complex infrared morphology and of its gas reservoir. There is growing evidence against the single, major wet merger scenario initially developed by \citet{webb_extreme_2015}. Although a cooling flow fits well with our constraints on the star-forming regions, X-ray data are needed to distinguish between this scenario and the gas stripping of multiples galaxies.}

%\item{We found an offset of 6$\sigma$ between our new VLA data and the SED fit previously made \citep{webb_extreme_2015}, which suggests that star-forming regions are not embedded in the BCG. Combining the SED fit with our upper limit for the radio emission induced by star formation, we find that star-forming regions are either made of numerous clumps or very extended.}

\end{description}

\section*{Acknowledgements}\label{ackno}
We acknowledge support from the NRAO helpdesk for data reduction bugs and from Tracy Clarke for persistent data reduction bugs. We wish to thank also Michael McDonald for sharing information about the Phoenix Cluster and Mark Voit for being the first to invoke the possibility of a cooling flow. AT is supported by the NSERC Postgraduate Scholarship-Doctoral Program. TMAW and JHL acknowledge the support of an NSERC Discovery Grant and of the FRQNT. JHL is also supported by NSERC through the Canada Research Chair programs. MM acknowledges support from the Spanish Juan de la Cierva program (IJCI-2015-23944). GW is supported by the National Science Foundation through grant AST-1517863, by \textit{HST} program number GO-15294, and by grant number 80NSSC17K0019 issued through the NASA Astrophysics Data Analysis Program (ADAP). Support for program numbers GO-13677/14327.01 and GO-15294 was provided by NASA through a grant from the Space Telescope Science Institute, which is operated by the Association of Universities for Research in Astronomy Incorporated, under NASA contract NAS5-26555. %\facility{VLA} %\facility{facility ID} or \facilities{facility ID, facility ID, facility ID, ... }
%\software{CASA} \citep{mcmullin_casa_2007} %\software{SExtractor \citep{1996A&AS..117..393B}

%\clearpage
%\bibliographystyle{authordate1}
%\bibliographystyle{aasjournal}
%\section*{References}
\bibliographystyle{mnras}
%\bibliography{refs} 
\bibliography{References}  

\bsp	% typesetting comment
\label{lastpage}
\end{document}